\documentstyle[12pt,epsf]{article}

\input{psfig}

\newcommand{\nc}{\newcommand}
\def\frac#1#2{{\textstyle {#1 \over #2}}}

\nc{\beq}{\begin{equation}}
\nc{\eeq}{\end{equation}}
\nc{\beqa}{\begin{eqnarray}}
\nc{\eeqa}{\end{eqnarray}}
\nc{\lsim}{\begin{array}{c}\,\sim\vspace{-21pt}\\< \end{array}}
\nc{\gsim}{\begin{array}{c}\sim\vspace{-21pt}\\> \end{array}}

\begin{document}
\begin{titlepage}

\begin{center} 

\vskip .5 in
{\large \bf 
The $2D$ $J_1-J_2$ XY and XY-Ising Models}

\vskip .6 in

  {\bf  P. Simon}\footnote{simon@lpthe.jussieu.fr }
   \vskip 0.3 cm
 {\it   Laboratoire de Physique Th\'eorique et Hautes Energies  }\footnote{ 
 Unit\'e associ\'ee au CNRS URA 280}\\
 {\it  Universit\'es Pierre et Marie Curie Paris VI et Denis Diderot Paris VII}\\
{\it  2 pl. Jussieu, 75251 Paris cedex 05 }

  \vskip  1cm   
\end{center}
\vskip .5 in
\begin{abstract}
We consider the $2D$ $J_1-J_2$ classical XY model on a square lattice. In the frustrated phase corresponding to $J_2>J_1/2$, an Ising order parameter emerges by an ``order due to disorder'' effect. This leads to a discrete symmetry plus the $O(2)$ global one. We formulate the problem in a Coulomb gas language and show by a renormalization group analysis that only two phases are still possible: a locked phase at low temperature and a disordered one at high temperature. The transition is characterized by the loss of Ising  and XY order at the same point. This analysis suggests that the $2D$ $J_1-J_2$ XY model is in the same universality class than XY-Ising models. 
\end{abstract}

 \vskip  1cm 
PACS NUMBERS: 05.70.+q, 75.10.Hk
\end{titlepage}
\renewcommand{\thepage}{\arabic{page}}
\setcounter{page}{1}
The ground state of a large class of two-dimensional classical XY models have the particularity to exhibit both continuous and discrete $Z_2$ degeneracy simultaneously in the ground state. It results   the appearence of a new Ising-like order parameter, in addition to the continuous $U(1)$ symmetry. It is the case of the fully frustrated XY models (FFXY) \cite{ffxy}-\cite{olson}, triangular lattice frustrated XY model (TFXY) \cite{trian}, helical XY models \cite{hxy}, Josephson junctions arrays in a transverse magnetic field at half flux per plaquet \cite{joseph}. The interplay between both order parameters can lead to a new type of unusual critical behavior, caracterized by the loss of Ising and XY order at the same point \cite{kos1,kos2}. Nevertheless, this scenario is still under controversy: some recent results of Olsson seem to indicate two very close transitions of Ising and Kosterlitz-Thouless type  \cite{olson} or a single but decoupled transition.\\\noindent
All this class of models with both continuous and discrete degeneracy are expected to be described by the  XY-Ising model consisting of Ising and XY models coupled each other \cite{kos2,kos3} defined by the following Hamiltonian
\beq
\label{isxy}
H_{XY_I}=-\sum\limits_{<i,j>}[(A+B\sigma_i\sigma_j)\cos(\theta_i-\theta_j)+C \sigma_i\sigma_j]~,
\eeq
\noindent
where $\sigma=\pm1$ is the Ising spin, $\theta$ the phase of a two component XY unit vector (XY spin) and $<~,~>$ indicates nearest neighbors. The above Hamiltonian has a very rich phase structure  with in particular, in the  $A=B$ plane, a branch corresponding to simultaneous loss of Ising and XY order \cite{kos2,kos1}. The critical behavior of the FFXY and TFXY seem to correspond to special points of this branch.\\\noindent

Nevertheless, there are some cases where the discrete $Z_2$ degeneracy can result of a continuous one which have been lifted by thermal or quantum fluctuations. It constitutes what is currently called an ``order by disorder'' effect \cite{vil1,hen}, in so far as fluctuations reduce possible ground states. Despite the basic symmetries become identical, once fluctuations have been incorporated, the relations with XY-Ising model described by (\ref{isxy}) do not seem  {\it a priori} obvious. Such a situation can be encountered in spins systems with competing interactions. The simplest possible model displaying all these characteristics is the $2D~J_1-J_2$ XY model on a square lattice whose Hamiltonian reads
\beq
\label{ham}
H= -J_1\sum\limits_{<i,j>}\cos(\theta_i-\theta_j)+J_2 \sum\limits_{<<k,l>>}\cos(\theta_k-\theta_l)~,
\eeq
\noindent
with $J_1,J_2>0$, and $<<~<>>$ indicates the sum over next to nearest neighbors. 
  When $2J_1>J_2$ the ground state is ferromagnetic. It leads to a Kosterlitz-Thouless (KT) transition at the temperature $T_{KT}={\pi(J_1-2J_2)\over 2}$ \cite{sim}. However, when $2J_1<J_2$, the ground state consists of two independent $\sqrt{2}\times\sqrt{2}$ sublattices with AF order. The ground state energy $E_0=-2NJ_2$ does not depend on $\phi$, an angle parametrizing the relative orientations of both sublattices. This non-trivial degeneracy is lifted by thermal fluctuations and a collinear ordering (corresponding to $\phi=0$ or $\pi$) is selectionned \cite{hen}. The angle $\phi$ thus plays a role analogous to an Ising  order parameter. 
 Monte-Carlo simulations predict at low temperature a phase with nematic order and a disordered phase at high temperature \cite{hen,fernan}. A first treatment which decouples both order parameters has been done in \cite{sim} suggesting that the expected
KT transition is lost and replaced by another one where the Ising order parameter plays an important role. \\\noindent
In this letter, we want to clarify the relations between the $J_1-J_2$ XY model and other models with $U(1)\times Z_2$ symmetry. We show that the Coulomb gas formulation of the frustrated phase coincides with the one of XY-Ising models when both sublattices are decoupled {\it i.e.} in the infrared regime. This result suggests that  the $J_1-J_2$ XY model belongs to the   class of XY-Ising models. Moreover, we give an estimate of the curve separating the locked and disordered phase in the (${J_2\over J_1},T$) plane.

\vskip 0.5cm
\noindent
We consider the ground state where both sublattices have independent AF order.
 The first step, following Chandra et al. \cite{chan}, is to perform a gradient expansion of the classical energy (\ref{ham}). The problem is now translated in a new one on a $(2\times 2)$ square lattice, but now with two spins $1$ and $2$ per vertices pointing in the same directions. The new classical action ${\cal A}$ reads
\beq
\label{action}
{\cal A} =-{2J_2\over 2T}\sum\limits_r \left[
(\vec{\nabla}\theta_1)^2+ (\vec{\nabla}\theta_2)^2 
+2\lambda\cos\phi~ (\nabla^x\theta_1\nabla^x\theta_2-\nabla^y\theta_1\nabla^y\theta_2)\right] ~,
\eeq
\noindent
where we have defined $\lambda={J_1\over 2J_2}<1$ and introduced the lattice derivatives $\nabla^x,~\nabla^y$ \cite{sim}. The signature of the $O(2)$ degeneracy lies now in the strong anisotropy between $x$ and $y$ directions. The $\cos\phi$ labels the different possible classical ground states at $T=0$. If we do the Gaussian integration, we recover the result of Henley \cite{hen}, namely $
{\cal A}\sim const-0.32\left({J_1\cos\phi\over 2J_2}\right)^2$
proving that a collinear ordering (with $\cos\phi=\pm1$) is effectively selected. 
Let us consider the action (\ref{action}) with $\cos\phi=1$ for example (the case $\cos\phi=-1$ can be deduced by changing $\lambda\to-\lambda$). We want to include vortex excitations in the spin wave action  (\ref{action}). The usual strategy is to apply the Villain transformation on each quadratic terms \cite{vil2} i.e. we introduce on each lattice links a gauge field in order the action (\ref{action}) to be $2\pi$ periodic. The drawbacks linked to the Villain approximation are the decoupling between spin waves and vortices. In this model, spin waves are responsible of the Ising order parameter appearence. The study of the models quoted above with both discrete and continuous degeneracy have proved this coupling to be strongly relevant \cite{ffxy}-\cite{kos2}. To cure this drawback, we introduce, following Chandra et al. \cite{chan}, a quadrupole coupling term $A_c\sim-0.32 \lambda^2\sum\limits_{\vec{r}}\cos^2(\theta_1(\vec{r})-\theta_2(\vec{r}))$, in the action
 (\ref{action}). This term just corresponds to local spin waves effects. Hence, this quadrupole coupling term plays the role of a symmetry breaking field which  can be treated as follow:
\beq
\exp\left[h\cos p (\theta_1(\vec{r})-\theta_2(\vec{r}))\right]=\sum\limits_{S(\vec{r})}\exp \left[ipS(\vec{r}
) (\theta_1(\vec{r})-\theta_2(\vec{r}))+\log y_sS^2(\vec{r})\right],
\eeq
\noindent
where in our case $y_s=h/2=0.08\lambda^2$ and $p=2$.
 To study the action  (\ref{action}), plus the symmetry breaking field, we first diagonalize the bilinear form in $\theta_i$ and then obtain two decoupled actions ${\cal A}_1$ and ${\cal A}_2$, where the $2\pi$ periodicity is introduced in quadratic terms. The partition function thus reads
\beq
\label{fpart}
{\cal Z}= \sum\limits_{\{n_1^{\mu}(r),l_1^{\mu}(r)\}}
 \sum\limits_{\{n_2^{\mu}(r),l_2^{\mu}(r)\}} \sum\limits_{S(r)}
\int {\cal D}\theta_1 {\cal D}\theta_2 e^{{\cal A}_1+{\cal A}_2}~,
\eeq
with,
\beqa
\label{action1}
{\cal A}_1&=&-{J_2\over 2T}\sum\limits_r \left[(\nabla^{\mu}\theta_1(r)-2\pi n_1^{\mu}(r))^2
+\lambda[(\nabla^1\theta_1(r)-2\pi l_1^1(r))^2- (\nabla^2\theta_1(r)-2\pi l_1^2(r))^2]
\right] \\
\label{action2}
{\cal A}_2&=&-{J_2\over 2T}\sum\limits_r \left[(\nabla^{\mu}\theta_2(r)-2\pi n_2^{\mu}(r))^2
-\lambda[(\nabla^1 \theta_2(r)-2\pi l_2^1(r))^2- (\nabla^2\theta_2(r)-2\pi l_2^2(r))^2]
\right]\nonumber\\
&&~~~~~~~~~~~~~~~~~~~~~~~~~~~~~~~~~~~~~~~~~~~~~~~~~+ip\sum\limits_r S(r)\theta_2(r) +\log y_s S^2(r)~.
\eeqa
\noindent
We have introduced for each action (\ref{action1}) and (\ref{action2}) four link variables per vertices $n_1^{\mu},l_1^{\mu}$ and $n_2^{\mu},l_2^{\mu}$ with $\mu=1$ for $x$ direction and $\mu=2$ for $y$ direction. The $n$ link variables correspond to AF interactions (proportional to $J_2$), the $l$ link variables to ferromagnetic interactions 
(proportional to $J_1$). It is equivalent to define two covariant  derivatives on the original lattice, one for unit edges, one for diagonal edges. Moreover, it enables to keep traces of the original lattice structure, and especially of all topological excitations it can support. \\\noindent
The action  (\ref{action1}) can be infered from (\ref{action2}) by taking $p=0$ and $\lambda\to-\lambda$. Because of the $2\pi$ periodicity, the four link variables in $ (\ref{action2})$ are not independent. It still remains three degree of freedom per sites plus the variabla $S(r)$. Following \cite{sim}, we introduce three integer valued variables $M_1(r)=\epsilon^{\mu\nu}\nabla^{\mu}n_2^{\nu}$, $M_2(r)=\epsilon^{\mu\nu}\nabla^{\mu}l_2^{\nu}$, 
$M_3(r)=\nabla^1 l_2^2-\nabla^2 n_2^1$ (with $\epsilon^{\mu\nu}$ the totally antisymmetric tensor), which represent three independent vortex variables. On the original lattice, $M_1(r),~M_2(r)$ are associated respectively, to the circulation of the gauge field $n$ around a $\sqrt{2}\times\sqrt{2}$ plaquet and of $l$ around a $1\times1$ plaquet. $M_3(r)$ is a mixture of $n$ and $l$ gauge variables around a $1\times \sqrt{2}$ plaquet. We suppose for convenience $\lambda<<1$, it will not change the results. After some tedious but standard manipulations, the action   (\ref{action2}) can be written as
\eject
\beqa
\label{vaction}
{\cal A}_2&=& \sum\limits_{r\neq r'}\left[\pi\beta(1+\lambda)M_1(r)\log {||r-r'||\over a}M_1(r')+\beta\lambda M_2(r)\log {||r-r'||\over a} M_2(r')\right.\nonumber\\
&&+2\beta\lambda M_1(r) \log {||r-r'||\over a} M_2(r')-2\beta\lambda(M_1(r)+M_2(r))
 \log {||r-r'||\over a} M_3(r')\nonumber\\
&&-ip(1-\lambda) M_1(r) \Theta||r-r'|| S(r')-ip\lambda M_2(r)\Theta||r-r'||S(r')\\
&&\left.+2ip M_3(r)  \Theta||r-r'|| S(r')+{p^2\over 4\pi\beta} S(r)\log {||r-r'||\over a}S(r')
\right]\nonumber\\
&&+\sum\limits_r\left[ \log y_1~ M_1^2(r)+ \log y_2~ M_2^2(r)+ \log y_3~ M_3^2(r)+\log y_s~ S^2(r)\right]~,\nonumber
\eeqa
\noindent
with $\beta={1\over T}$. In this equation, we have defined a special norm $||r||^2= {x^2\over 1-\lambda}+{y^2\over 1+\lambda}$ due to the anisotropy. In a similar way, $\Theta||r-r'||=\arctan({y-y'\over x-x'}\sqrt{{1-\lambda\over 1+\lambda}})$. We have also included four fugacities defined initially  by $y_1=\exp(-\pi^2\beta(1+\lambda)/2)$, $y_2= \exp(-\pi^2\beta\lambda/2)$, $y_3=1,~y_s= 0.08\lambda^2\exp(- p^2/8\beta)$ associated to the three vortices and to $S(r)$. They can also be regarded as genuine variables. In order to obtain the formula (\ref{vaction}), we have neglected terms with higher power of $\lambda$. The coupling terms between $S(r)$ and vortices variables can be obtained following ref. \cite{nel} and using $\lambda<<1$.\\\noindent
Notice that $M_3(r)$ does not couple with itself, independently of the approximation $\lambda<<1$. This means that vortices $M_3(r)$ are already present in the collinear ground state we consider initially. To understand the role of vortices $M_3$, one can study the action   (\ref{vaction}) with $p=0$ and $\lambda\to-\lambda$, in the region $y_k<<1$ in order to restrict the charges to take only the values $0$ and $\pm1$. A straightforward generalization of Kosterlitz-Thouless equations \cite{sim} proves that the flow associated to vortices $M_2$ and $M_3$ is driven toward a high density regime as could have been thought (see ref. \cite{sim} for details). Moreover $\lambda={J_1\over 2J_2}$ runs quicky toward zero i.e the two sublattices decouple, as in the case of Heisenberg spins \cite{chan}. This justifies the approximation $\lambda<<1$. A similar phenomena is naturally observed for the flow associated to the more general action (\ref{vaction}). How could we interprete this ?
In \cite{sim}, it was argued that this high density regime for vortices $M_2$ and $M_3$ is responsible of the destabilization of the expected KT fixed point associated to $M_1$ vortices. Following this line, one can go one step further. The KT flow indicates a proliferation of $M_2$ and $M_3$ vortices {\it i.e.} they try  to fill the semi-classical vacuum, in other words they condense, so impose geometrical constraints on the ground state.
Had we initially considered a ground state with independent AF order on each sublattices (so $\phi$ as a parameter in (\ref{action}) ), the KT equations would tell us that vortices $M_2$ and $M_3$  condense and  $\phi=0$ or $\pi$ are the only configurations satisfying both condensate of vortices. Consequently, the collinear ground state can also be regarded as a lattice of vortex. 
  This implies that the degrees of freedom associated to $M_2$ and $M_3$ (so to $l^{\mu}$)   quickly freeze in the infrared limit. If the degrees of freedom associated to $l^{\mu}(\vec{r})$ are frozen, the remaining excitations are (see (\ref{vaction})) vortices excitations on both diagonal sublattices ($M_1$) and local spin wave effects  {\it i.e.} the symmetry breaking field, which remains relevant even in the regime $\lambda\to 0$ (see ref. \cite{jkkn}). In this infrared  regime the action ${\cal A}={\cal A}_1+{\cal A}_2$  considerably simplifies and an effective action ${\cal A}_{eff}$ can be drawn (where we have recoupled both actions and rescaled  $\beta$)
\beqa
\label{actfin}
{\cal A}_{eff}&=&\sum\limits_{r\neq r'}\left[\pi\beta  J_2 M_1(r) \log{|r-r'|\over a} M_1(r')
+\pi\beta' J_2 M'_1(r) \log{|r-r'|\over a} M'_1(r')\right.\nonumber\\
&&-ip (M_1(r)+M'_1(r))  \Theta|r-r'| S(r')
+\left.{p^2\over 2\pi\beta} S(r)\log {|r-r'|\over a}S(r')
\right]\\&&+\sum\limits_r\left[ \log y_1~ (M_1)^2(r)+\log y'_1~ (M'_1)^2(r)+\log y_s~ S^2(r)\right]~\nonumber,
\eeqa
where $\beta=\beta'$, and $M_1$ and $M'_1$ are vortices excitations on both diagonal sublattices and $y_1,~y'_1$ their associated fugacities. The effective action (\ref{actfin}) 
corresponds to two coupled XY models. Under renormalization, the condition $\beta=\beta'$ is preserved and the coupling term $h$ is strongly relevant and locks the phase difference
 in $\theta_1(r)=\theta_2(r)+k\pi$ with $k=0,1$ \cite{doma,kos3}. It leads in the strong coupling limit ($h>>1$) to an effective XY-Ising model (\ref{isxy}) with $A=B$ and $C$  depending on initial values of $h$ and $\beta$  \cite{kos3,kos2}. It follows from this analysis that the critical behavior of the $J_1-J_2$ XY model is described by a coupled doubled Coulomb gas which is in the class of XY-Ising models.  Hence, this suggests only
one transition (or two very closed transitions) with simultaneous loss of Ising and XY order in accordance with Monte Carlo simulations \cite{hen,fernan}.  It is worth while noting that  the $J_1-J_2$ XY model enters in the universality class  of XY-Ising model in a non-trivial way, by dynamical spin wave effects contrary to other models quoted in the introduction.
 The line separating the locked from the unlocked phase  depend on the initial conditions for the fugacities. This line is caracterized approximately by $ y_1,y_1'\sim y_s$ initially. It enables us to  estimate the critical temperature $T_c/J_2$ for a given
value of $\lambda=J_1/2J_2$:
 \beq
{T_c\over J_2}\sim0.5*[\log(\gamma\lambda^2)+ \sqrt{\log^2(\gamma\lambda^2)+2\pi^2}]
\eeq
The associated phase diagram has been represented in Figure 1.
For $J_2=J_1$, we find $T_c/J_2=1.0$ in accordance with the results of Henley \cite{hen} $T_c/J_2=0.97\pm 0.02$ and of Fern\'andez {\it et al.} \cite{fernan}, $T_c/J_2=0.9\pm 0.02$. Yet, no definitive conclusion can be drawn because of the lack of numerical result.
In the limit $J_1/J_2\to 0$, the critical temperature converges logarithmically toward $0$ as it has already been checked for quantum Heisenberg spins \cite{chan}.\\\noindent
The phase diagram of the XY-Ising model (\ref{isxy}) consists of three branches which meet at a multicritical point. One of the branches separates the locked from the unlocked phase whereas the other two separate KT and Ising transitions. 
The models quoted in the introduction, where only one transition was observed seem to correspond to special points on this  branch. Contrary to the fully frustrated XY model or triangular antiferromagnets, we have one more parameter in the $J_1-J_2$ XY model,
therefore a line of transition in the $(J_1/J_2,T)$ plane.  When $J_2>>J_1$, our Coulomb gas representation (\ref{vaction}) represents two XY models very weakly coupled (the central charge will be close to 2) and it is plausible to think that this limit corresponds to   the tricritical point in the XY-Ising phase diagram where the critical line becomes first order. As we increase $J_1$, we will go away from the  tricritical point in direction of the multicritical point where the three branches meet. So,
we expect that the whole transition line in the $(J_2/J_1,T)$ plane corresponds to a part of the critical line in the $(A,C)$ plane. A similar situation was recently conjectured for
a model close to FFXY model \cite{benak}.

To sum up, we have shown using effective actions, that the $J_1-J_2$ XY model on a square lattice is in the same universality class of the XY-Ising model. The transition seems to correspond to a whole part (between the branch and tricritical points) of the critical line in the phase diagram of the XY-Ising model. This could be used to
test numerically the continuous variations of the central charge and critical exponents
along this critical line.

\vskip .5 in
{\bf Acknowledgements}\\\noindent
I would like to thank B. Dou\c{c}ot for stimulating discussions.

\eject
\begin{center}
{\bf Figure Caption}
\end{center}
FIG. 1 : The phase diagram associated to the  frustrated phase of the 2D, $J_1-J_2$ XY model. The transition between the disordered and the locked phases is characterized by a simultaneous loss of Ising and XY order like in XY-Ising models. Near the Lifshitz point, where the treatment fails, the curve has been extrapolated.

\psfig{figure=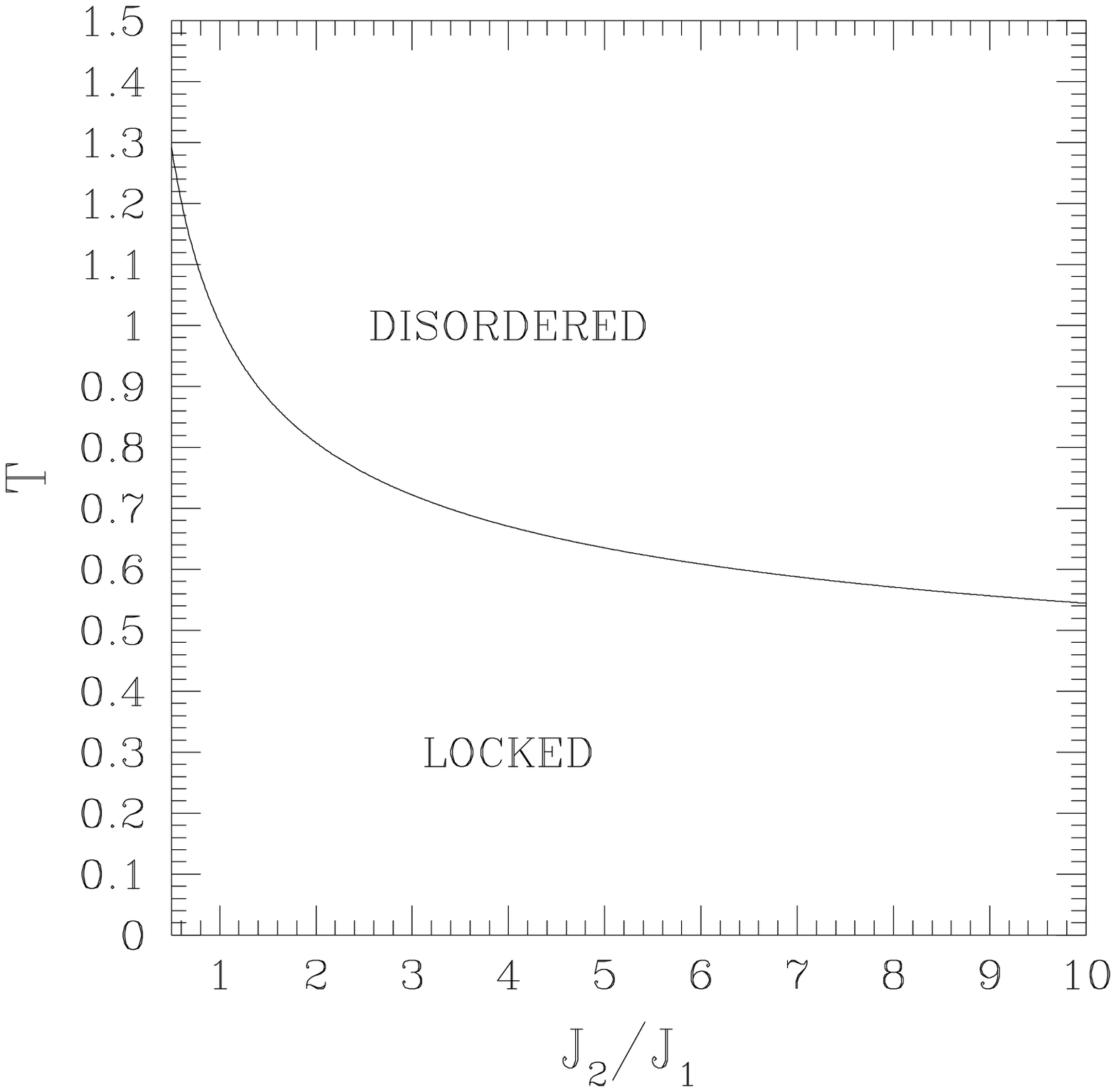,height=14cm,width=18cm}

\end{document}